\documentclass[review]{elsarticle}

\usepackage{lineno,hyperref}
\usepackage{graphicx}
\usepackage{subfig}
\usepackage{geometry}
\usepackage{booktabs}
\usepackage{multirow}
\usepackage{amssymb}
\usepackage{textcomp}
\usepackage{gensymb}
\usepackage{color, xcolor}
\hypersetup{pdfauthor=author}
\modulolinenumbers[5]

\journal{Nuclear Instruments and Methods in Physics Research Section A}

\begin{document}

\begin{frontmatter}

\title{\boldmath Cable Loop Calibration System for Jiangmen Underground Neutrino Observatory }

\author[1]{Yuanyuan Zhang}
\author[1]{Jiaqi Hui}
\author[1,2]{Jianglai Liu}
\author[3,4]{Mengjiao Xiao\corref{mycorrespondingauthor1}}
\cortext[mycorrespondingauthor1]{Corresponding author: mengjiaoxiao@gmail.com}
\author[1]{Tao Zhang\corref{mycorrespondingauthor}}
\cortext[mycorrespondingauthor]{Corresponding author: tzhang@sjtu.edu.cn}
\author[1]{Feiyang Zhang}
\author[1]{Yue Meng}
\author[2,1]{Donglian Xu}
\author[2]{Ziping Ye}

\address[1]{School of Physics and Astronomy, Shanghai Jiao Tong University, Shanghai Key Laboratory for Particle Physics and Cosmology, Shanghai 200240, China}
\address[2]{Tsung-Dao Lee Institute, Shanghai Jiao Tong University, Shanghai, 200240, China}
\address[3]{Department of Physics, University of Maryland, College Park, Maryland 20742, USA}
\address[4]{Center of High Energy Physics, Peking University, Beijing 100871, China}

\begin{abstract}
  A cable loop source calibration system is developed for the Jiangmen
  Underground Neutrino Observatory, a 20~kton spherical liquid
  scintillator neutrino experiment. This system is capable of
  deploying radioactive sources into different positions of the
  detector in a vertical plane with a few-cm position
  accuracy. The design and the performance of the prototype are
  reported in this paper.
\end{abstract}


\end{frontmatter}

\linenumbers

\section{Introduction}

The Jiangmen Underground Neutrino Observatory (JUNO) is a 20~kton
multi-purpose liquid scintillator (LS) detector, located 53~km from
the Yangjiang and Taishan nuclear power plants in south
China~\cite{yellow-book}. Its physics goal is to determine the
neutrino mass ordering (MO) via the measurement of electron
anti-neutrino energy spectrum with unprecedented
precision~\cite{petcov-original-paper,petcov2,learned,zhanliang-original-paper-2008}
and study the neutrino oscillation and the astrophysical
neutrinos~\cite{yellow-book}. The LS of the central detector (CD) is
confined by an acrylic sphere with an inner diameter of 35.7~m,
outside of which there is an ultra-pure water shielding. Approximately
$17000$ 20-inch photomultiplier tubes (PMTs) and $25000$ 3-inch PMTs
are located in the water viewing the LS.

Detector non-uniformity needs to be calibrated to ensure the energy
resolution required by the MO determination. Simulation studies have
shown that we need to deploy radioactive sources to about 200
locations in a vertical plane of the detector and to make use of the
azimuthal symmetry of the detector~\cite{ZFY_paper}. In this paper, we
discuss the design and the prototype of an automatic system which can
deliver the source to a wide range of positions via a cable loop, thus
named the Cable Loop System or CLS. The design concept has been inspired by similar systems in SNO~\cite{BOUDJEMLINE2010171},
KamLAND~\cite{Berger_2009}, and Borexino~\cite{Back_2012} experiments.

\begin{figure}
  \centering
  \includegraphics[width=0.8\textwidth]{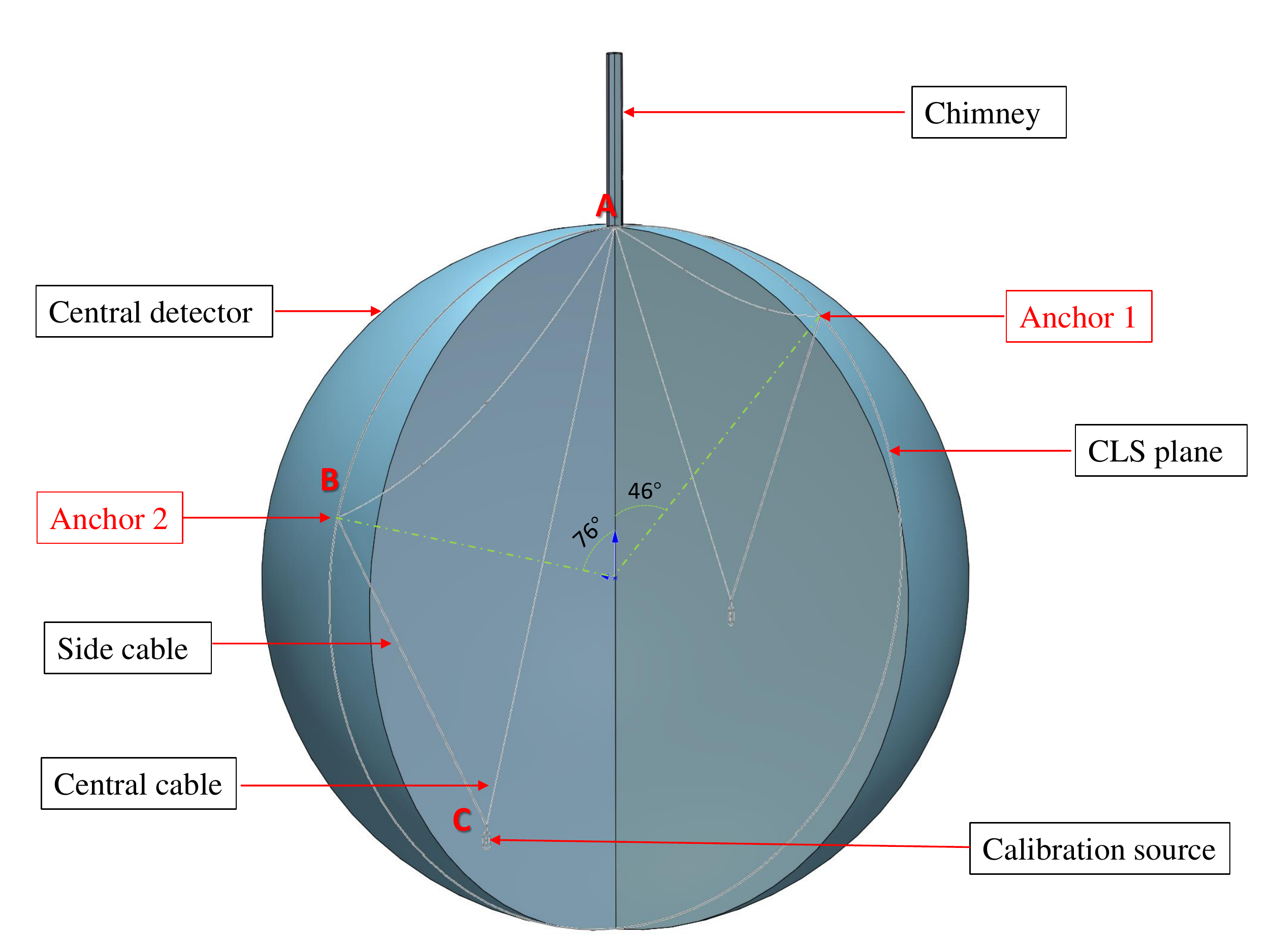}
  \caption{An illustration of the CLS. Two independent CLSs are
    envisioned to be installed in the JUNO detector to cover two half
    planes.}
  \label{CLS_total_concept}
\end{figure}

A sketch of the cable routing of the CLS is illustrated in
Fig.~\ref{CLS_total_concept}. The cable loop is made up by a
continuous cable which goes into the detector from the chimney, makes
a complete loop via A $\rightarrow$ B $\rightarrow$ C $\rightarrow$ A,
then runs out of the detector. Point C is the attachment point of the
source, on which an ultrasonic emitter is attached to allow accurate
positioning. The segments connecting A $\rightarrow$ B $\rightarrow$ C
and C $\rightarrow$ A are referred to as the side cable and the
central cable, respectively. The source can be delivered to a given
position by adjusting the lengths of the side cable and the central
cable. The JUNO detector is mostly azimuthally symmetric by design,
but the response function in radial and polar angle direction is
complex~\cite{ZFY_paper}. Therefore, we have to access the off-center
locations in a vertical half plane at the very least. In JUNO we
design two independent CLSs, each covering a half-plane, to reach a
larger accessible area and to facilitate the cross-check of the azimuthal
symmetry.

The rest of this paper is organized as follows. In
Section~\ref{sec:assembly}, we describe the design and prototype of
the CLS, followed by a discussion of the motion sequence in
Section~\ref{sec:calibration_strategy}. The performance of the
prototype and the projected accessible area for the full-size detector
are discussed in Section~\ref{sec:performance}, followed by a summary
in Section~\ref{sec:summary}.

\section{System design and prototype construction}
\label{sec:assembly}

\subsection{Overview of the CLS in the calibration house}

\begin{figure}[!htbp]
\centering
\includegraphics[width=4in]{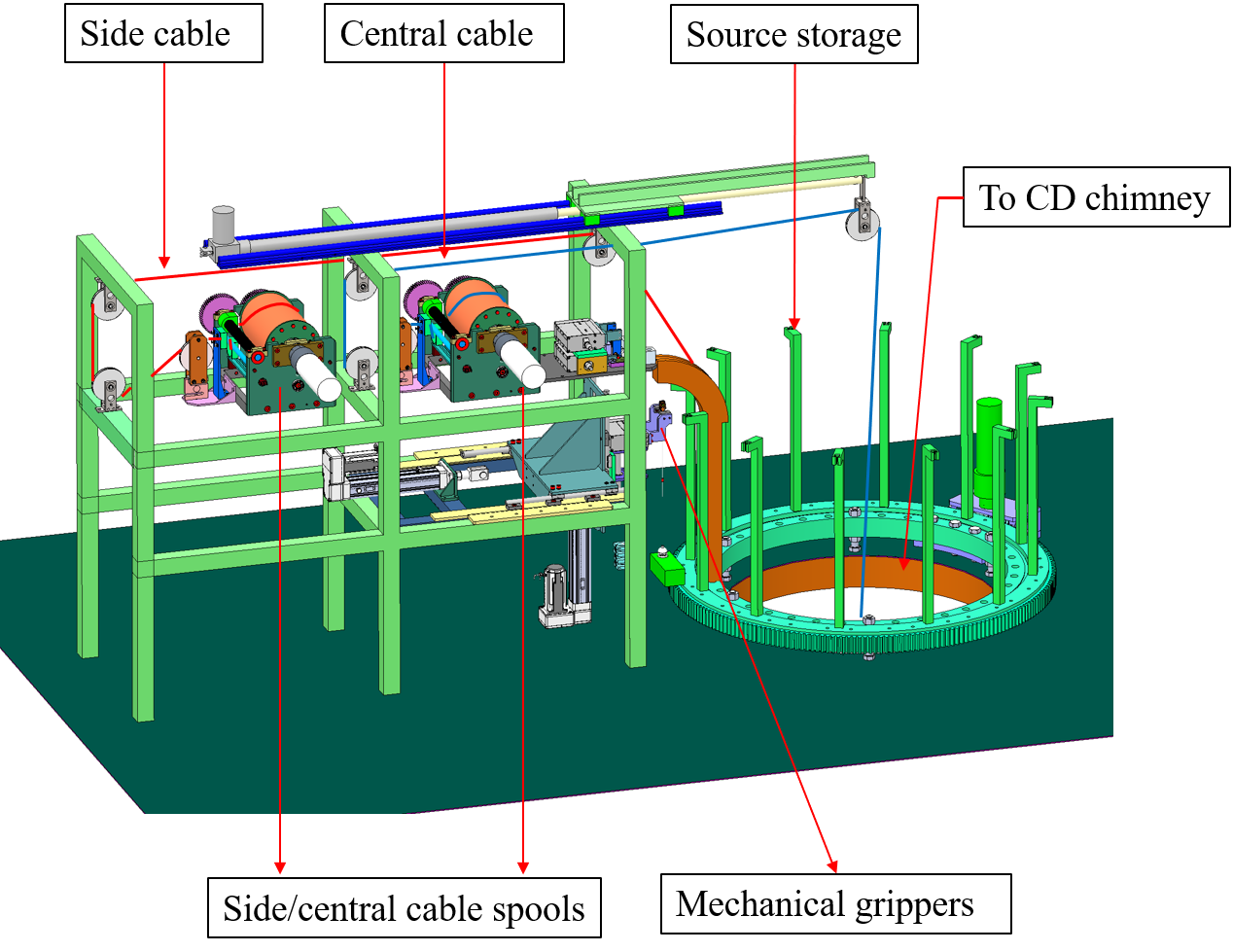}
\caption{Side view of one CLS in the calibration house. See text for details.}
\label{calibration_house}
\end{figure}

The mechanical source delivery systems are located in the calibration
house on top of the central detector, shown in
Fig.~\ref{calibration_house}. For each CLS, two spools are used to
store the cables and to drive the central and side cables
independently. Source assemblies are placed on the hanging frames on a
rotatable ring. A source can be attached to the CLS (point C in
Fig.~\ref{CLS_total_concept}) and restored automatically by mechanical
grippers or manually via the glove boxes.

\subsection{Components of the CLS}
\subsubsection{Stainless steel cables}

In the CLS, a large fraction of the cable will be permanently immersed
in the liquid scintillator. The cable should satisfy four
requirements: 1) be compatible with the LS in JUNO; 2) can supply
power to the ultrasonic emitter~\cite{2019Ultrasonic}; 3) be flexible
for source deployment and robust for more than 20 years of usage; 4)
do not introduce significant radioactive background.

\begin{figure}[!htb]
\centering
\includegraphics[width=3.5in]{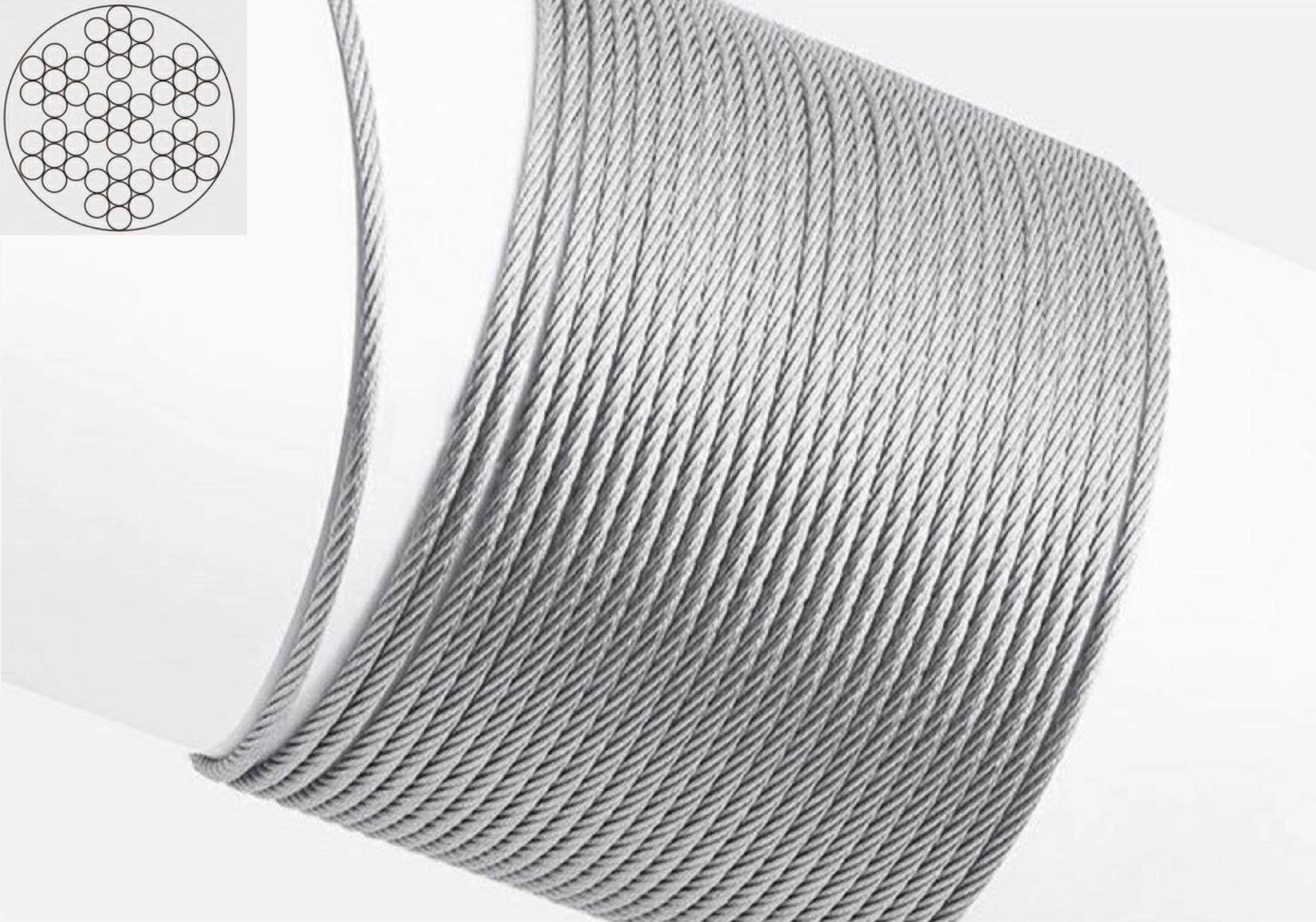}
\caption{A picture of the selected CLS cable. The top left corner
  shows the cross section of the cable.}
\label{SSwire}
\end{figure}

The selected cable is shown in Fig.~\ref{SSwire}. The Fluorinated
ethylene propylene (FEP) jacket with 0.1~mm thickness is compatible
with the LS, also with a low infiltration of the LS, which will bring little LS to the calibration house during the calibration. The cable has a 1.0~mm overall diameter with a turning
radius of less than 5~mm. Its core consists of $7\times7$ strands of
stainless steel wires, which makes that thermal expansion of the cable introduces a negligible effect. The average density of the cable is around
$3.565$~g/cm$^{3}$. The breaking strength is measured to be
60~kg. Each complete cable loop has a total length of 130~m,
approximately. The radioactivity of the selected cable is reported
elsewhere~\cite{ZFY_paper_background}, leading to negligible
contribution to the detector background.

\subsubsection{Chimney collar and CLS anchor}

\begin{figure}[!htb]
  \centering
  \subfloat[]
  {
    \begin{minipage}[t]{0.48\linewidth}
        \centering
        \includegraphics[width=1.\textwidth]{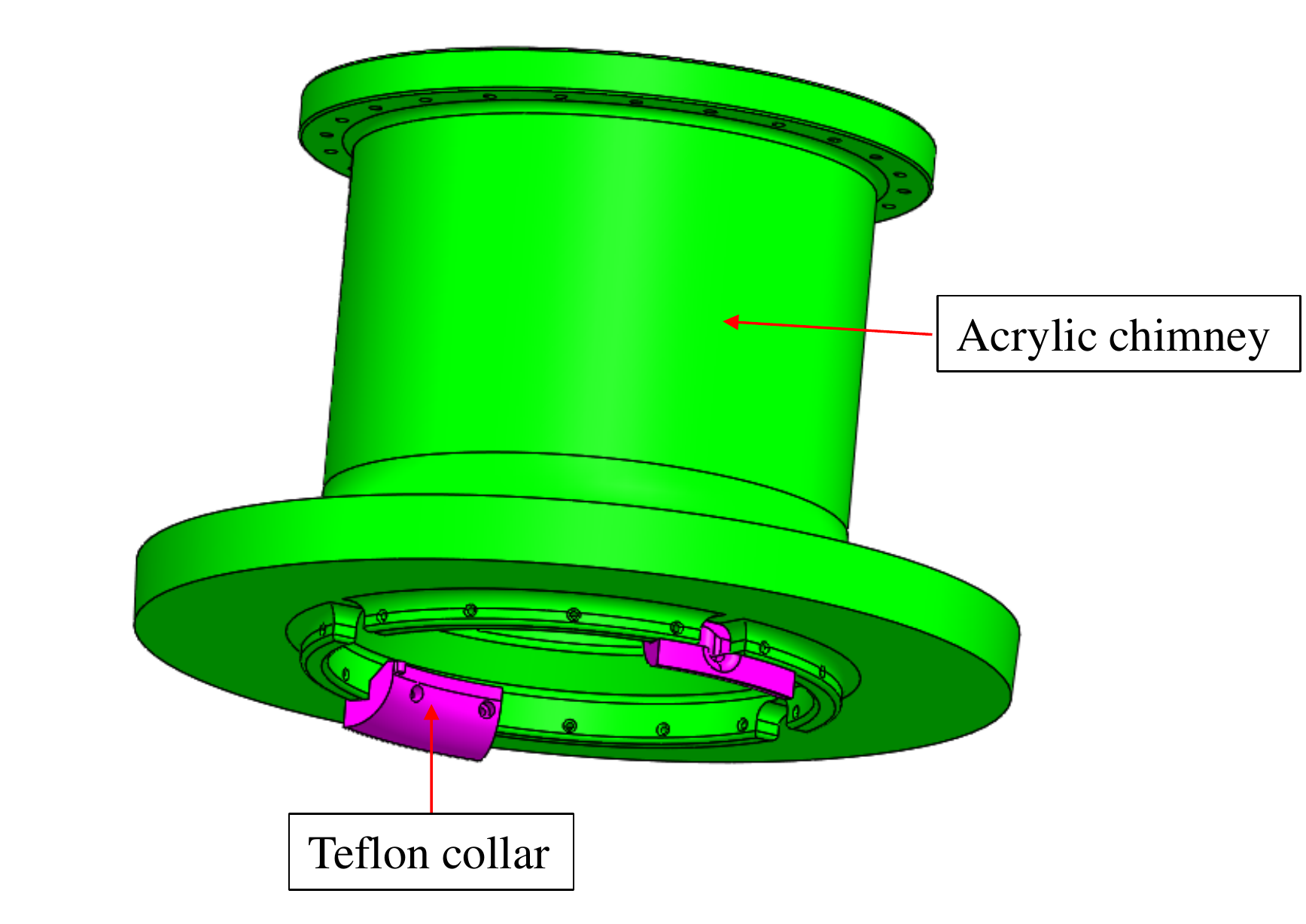}
        \label{structure_model1}
    \end{minipage}%
  }
  \subfloat[]
  {
    \begin{minipage}[t]{0.48\linewidth}
        \centering
        \includegraphics[width=1.\textwidth]{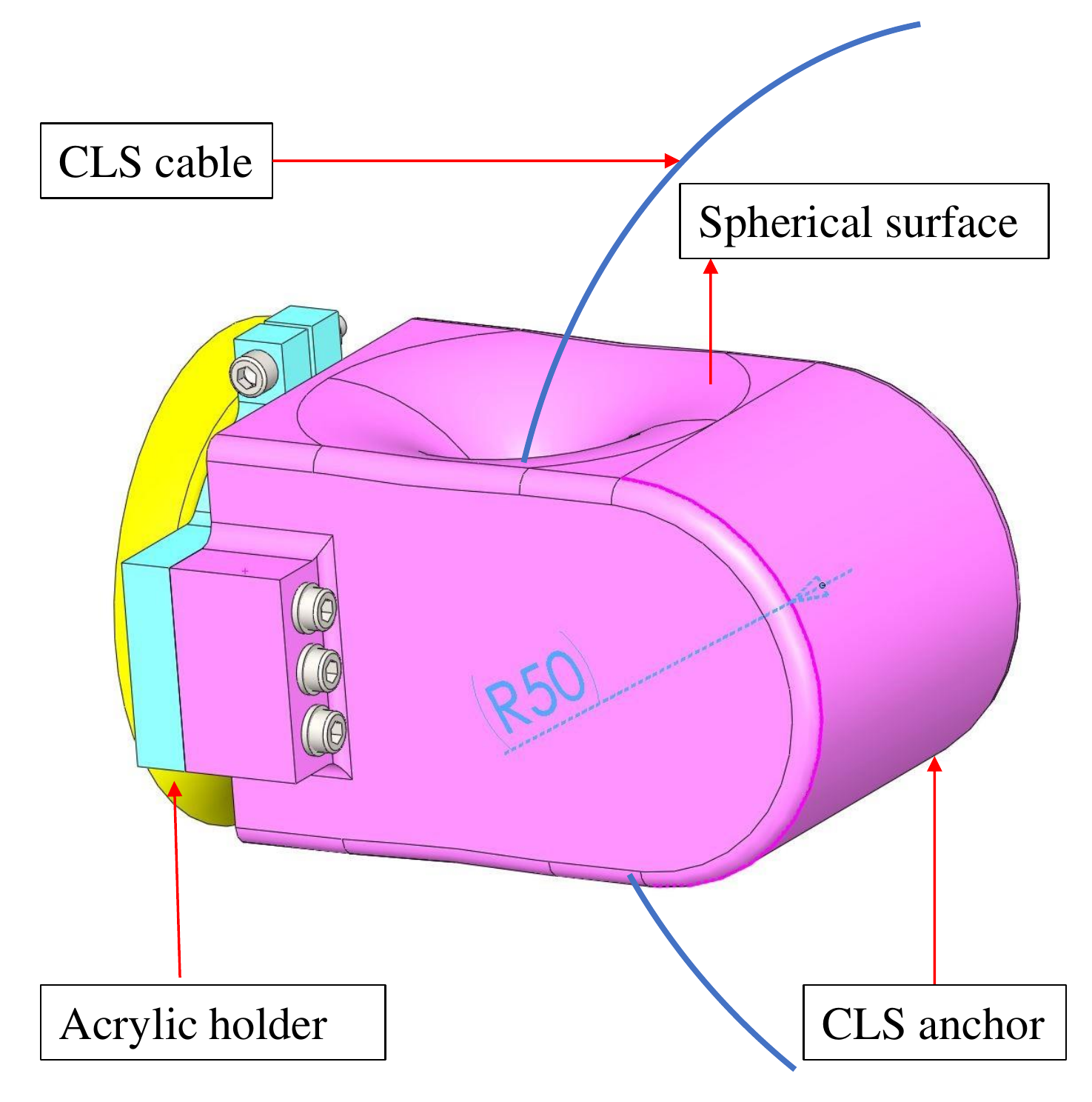}
        \label{structure_model2}
    \end{minipage}
  }%

  \caption{Design drawing of the chimney collars (a), mounted on the
    lowest segment of the CD chimney, and the CLS anchor with mounting
    fixture (b), the cable passes the anchor through a 1~mm hole.}
\end{figure}

In order to protect the joint between the chimney and the CD from the
abrasion of the cables, a Polytetrafluoroethylene (PTFE) collar will
be mounted on the underside of the joint, as shown in
Fig.~\ref{structure_model1}. The PTFE has a small friction coefficient
of 0.04, high photon reflectivity, low radioactivity, and is
compatible with the LS. Two identical PTFE cable anchors with a
spherical surface and a 50~mm turning radius (Fig.~\ref{structure_model2}) are designed to be
installed on the inner surface of the acrylic sphere at polar angles
$46\degree$ and $76\degree$, optimized by
simulations~\cite{ZFY_paper}. Despite that the frictions of these
designs are larger than those using pulleys, the robustness is our top
consideration for these components immersed permanently in the LS.

\subsubsection{Cable spools}

\begin{figure}[!htb]
  \centering
  \includegraphics[width=5in]{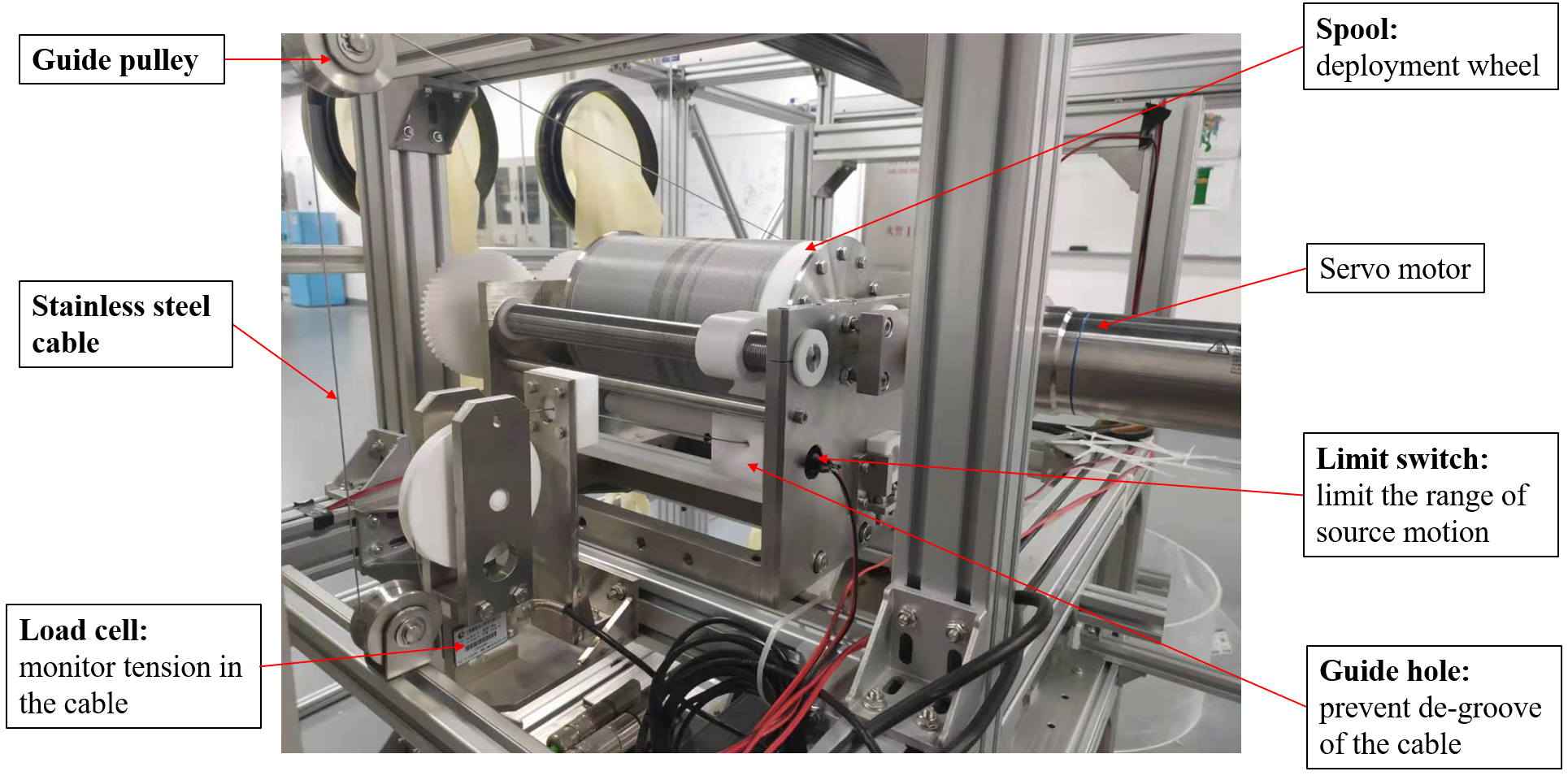}
  \caption{A picture of one cable spool of the CLS.}
  \label{winding_mechanism}
\end{figure}

Fig.~\ref{winding_mechanism} shows a picture of the cable spool system
for one CLS. Two identical cable spools are designed, each driving the
central/side cable. The spool is designed based on similar system used
in the Daya Bay automated calibration
system~\cite{Liu2013Automated, Zhang_2019}. The load cell with an accuracy of
0.025~N is used to monitor the tension in the cables. A cable
guide hole is designed to prevent de-groove of the cable. A
servo motor (AKMH22E-CNT2GE5K~\cite{servo_motor}) is used to drive the
spool, with its current monitored by the control software to avoid
excessive torque. In order to reduce the friction and protect the
stainless steel cables, all of the pulleys and spools are made out of the PTFE.

\subsubsection{Source assembly}

\begin{figure}[!htb]
  \centering
  \subfloat[]
  {
    \begin{minipage}[t]{0.45\linewidth}
        \centering
        \includegraphics[width=1.\textwidth]{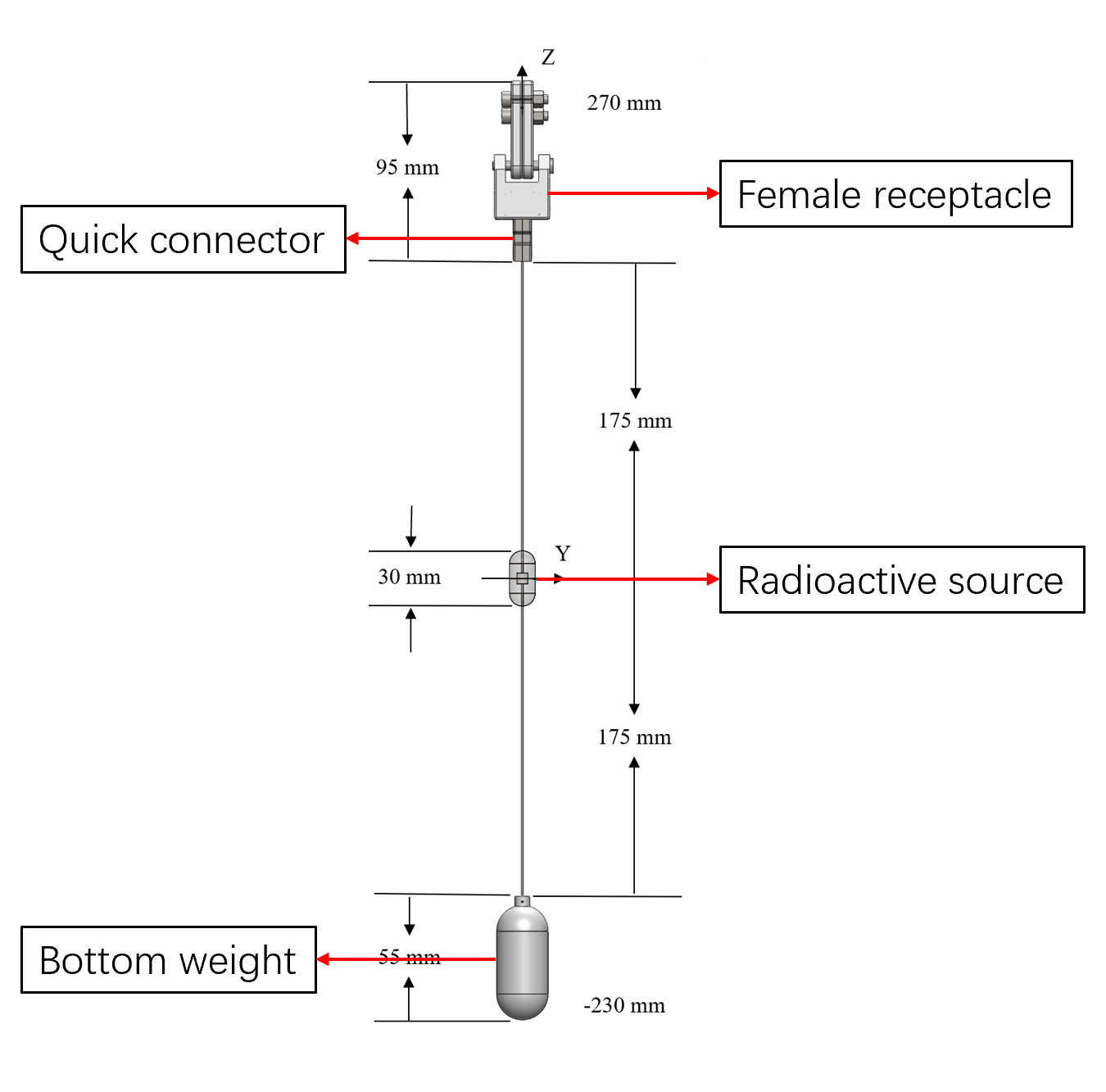}
        \label{source}
    \end{minipage}%
  }
  \subfloat[]
  {
    \begin{minipage}[t]{0.4\linewidth}
        \centering
        \includegraphics[width=1.\textwidth]{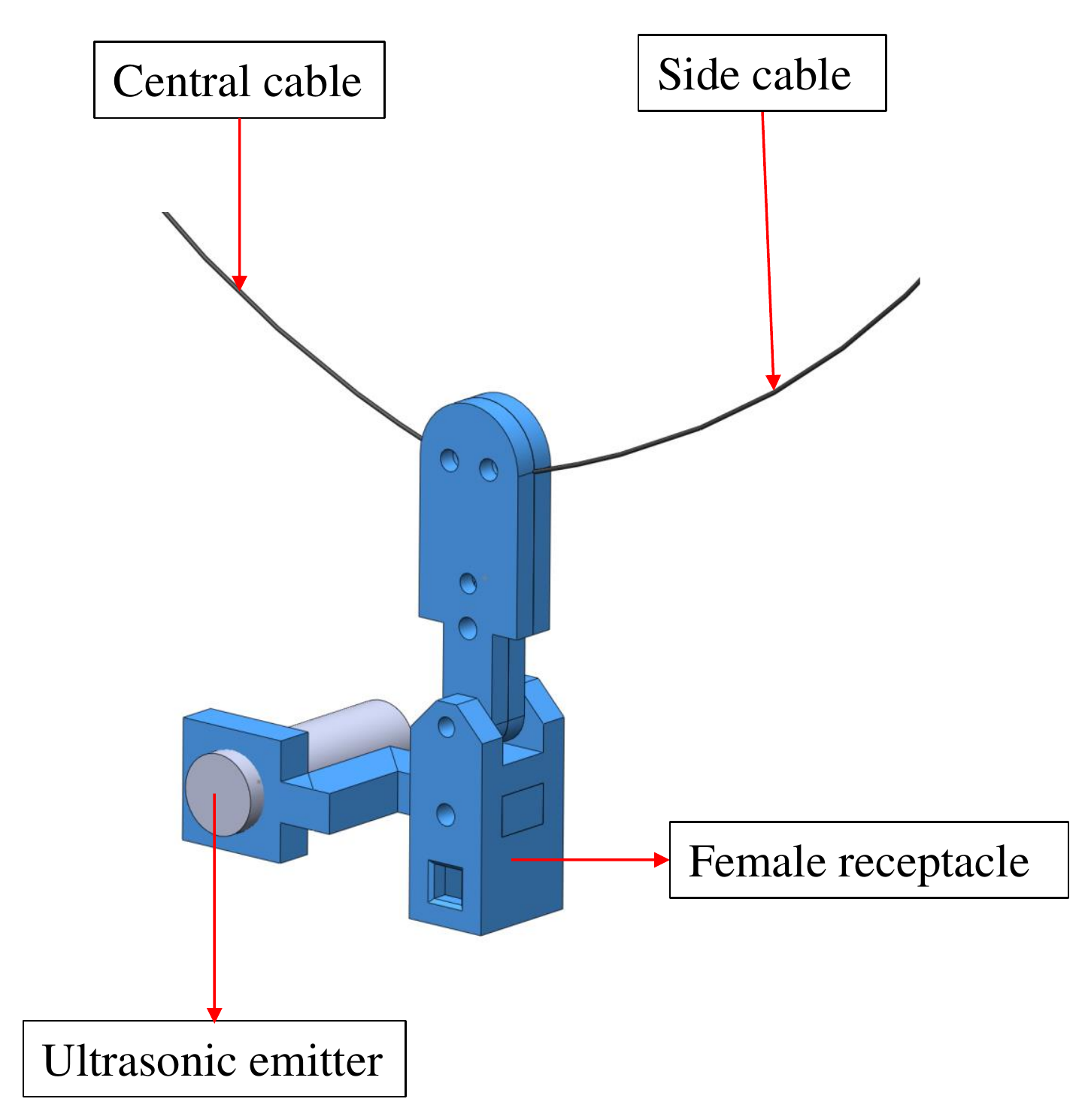}
        \label{female_plug}
    \end{minipage}
  }%

  \caption{Design drawing of a source assembly (a); details of a
    female receptacle and the ultrasonic emitter (b).}
\end{figure}

The basic design of a source assembly, from top to bottom, consists of
a quick connector, a radioactive source container, and a bottom
weight, inter-connected by about 20~cm of stainless steel cables, as
shown in the Fig.~\ref{source}. The quick connector (male plug) can be
inserted into the female receptacle attached to the CLS cable
(Fig.~\ref{female_plug}, also point ``C'' in
Fig.~\ref{CLS_total_concept}). The ultrasonic emitter is attached to
the female receptacle, with the CLS cable supplying electric driving
power for the emitter. The bottom weight provides additional
tension to keep the cables straight and to avoid cable slipping
grooves. The bottom weight ($\sim$ 50~g) is made with normal nickel to allow
potential rescue by a magnet. All pieces are made with PTFE
enclosures to optimize photon reflectivity. The total weight of the
source assembly (including the female receptacle and the ultrasonic
emitter) is approximately 150~g.

\newpage

\subsection{Risks and mitigations}

We have analyzed major risks associated with the failure of the CLS system, and incorporated mitigations into the system design. The
considerations are summarized in Table~\ref{table1}. Each risk is
counteracted with multiple mitigations.

\newpage

\begin{table}[!htbp]
   \caption{The possible failure modes in the CLS}
  \label{table1}
  \centering
  \begin{tabular}{|p{0.2\textwidth}|p{0.2\textwidth}|p{0.6\textwidth}|}
  \hline
  failure mode & risk  & mitigations\\
  \hline
  cable loses tension & cable slips groove on the spool & 1) set load cells lower limit to 0.2 N; 2) reverse motion to tighten the cable; 3) guide the cable back into grooves by hand via the glove box. \\
  \hline
  source assembly gets stuck (e.g. against the collar) & damage of the cable or source assembly & 1) smooth surfaces in the design; 2) set the load cell upper limit to 10 N; 3) reverse motion then slowly back up.\\
  \hline
  load cell failure & cannot monitor the state of cables & 1) use driving current in the servo motor as a tension indicator and set appropriate range; 2) visual feedbacks; 3) cover the chimney opening, access the calibration house, and replace the load cells when desired.\\
  \hline
  mechanical component failures & cannot move the source & 1) pull central cable by hand through the glove box to rescue the source; 2) cover the chimney opening, access the calibration house, and replace failed components when desired.\\
  \hline
  power off during the calibration &  calibration interruption & 1) use UPS system to leave enough time to exit the calibration gracefully; 2) use industrial PLC system for the motion control~\cite{PLC}, which keeps all data during power outage; 3) control software will save all data onto disk, which can be reloaded when restarted.\\
  \hline
  cable breakage & source drops into the CD & 1) set upper limits in load cells and current limits in servo motors so cables never experience tensions larger than 10 N; 2) load test all cables/joints to 10 times of the maximum allowable tension; 3) were the source dropped into the CD, fish the source with the ROV~\cite{NWPU-paper} carrying a magnet.\\
  \hline
  failure of the anchor de-attached from the acrylic sphere & anchor drops into the CD & 1) ensure the tension in the cable to be less than 20~N, corresponding to a conservative maximum local stress of 0.35~MPa (the long term allowable stress in acrylic is 5~MPa~\cite{Acrylic}); 2) attachment is fastened by bolt and nuts to avoid threads in the fixture\\
  \hline
\end{tabular}
\end{table}

\subsection{Prototype construction}

\begin{figure*}[!htbp]
  \centering
  \includegraphics[width=5in]{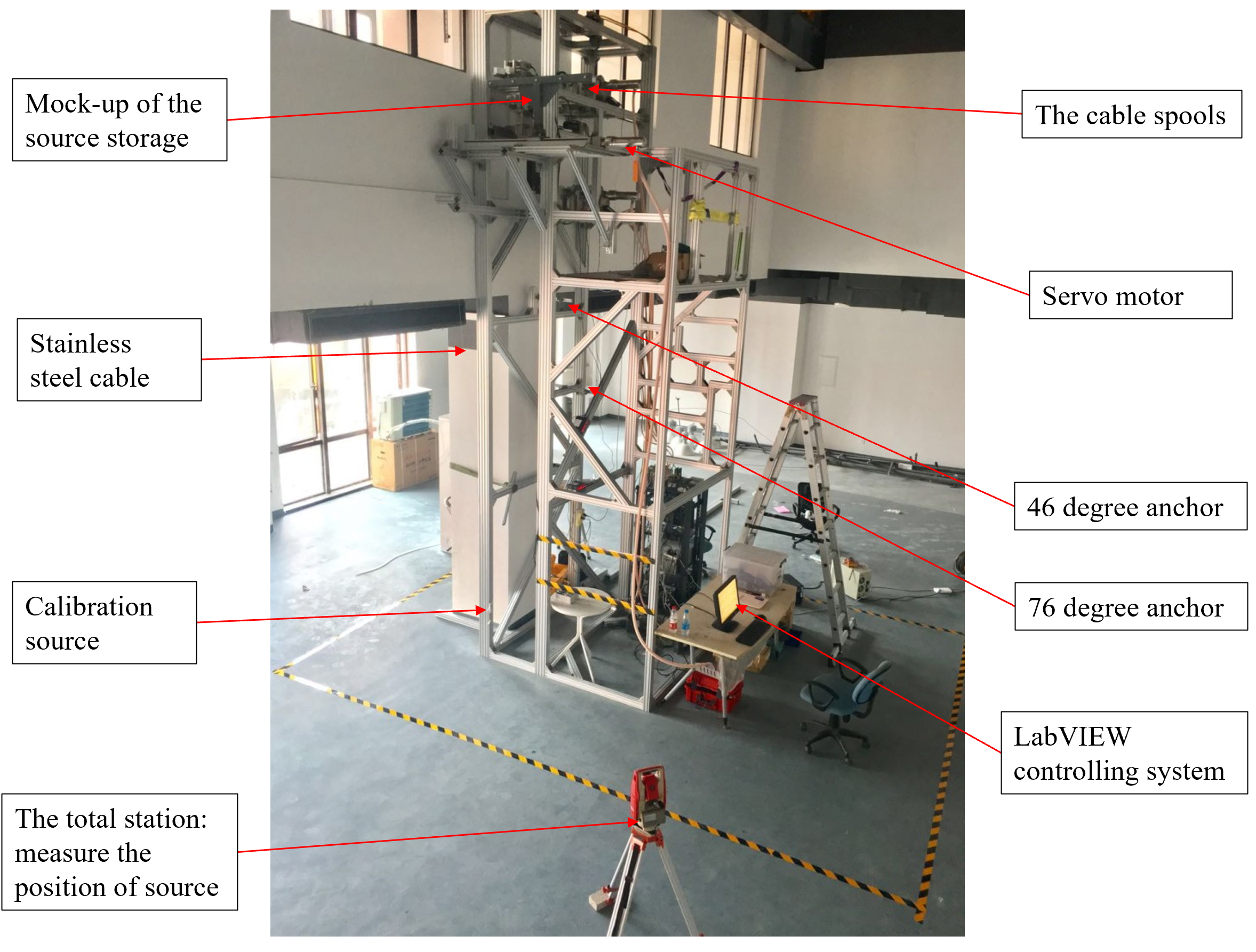}
  \caption{A picture of the CLS prototype in the lab.}
  \label{system}
\end{figure*}

\begin{figure*}[!htbp]
  \centering
  \includegraphics[width=5in]{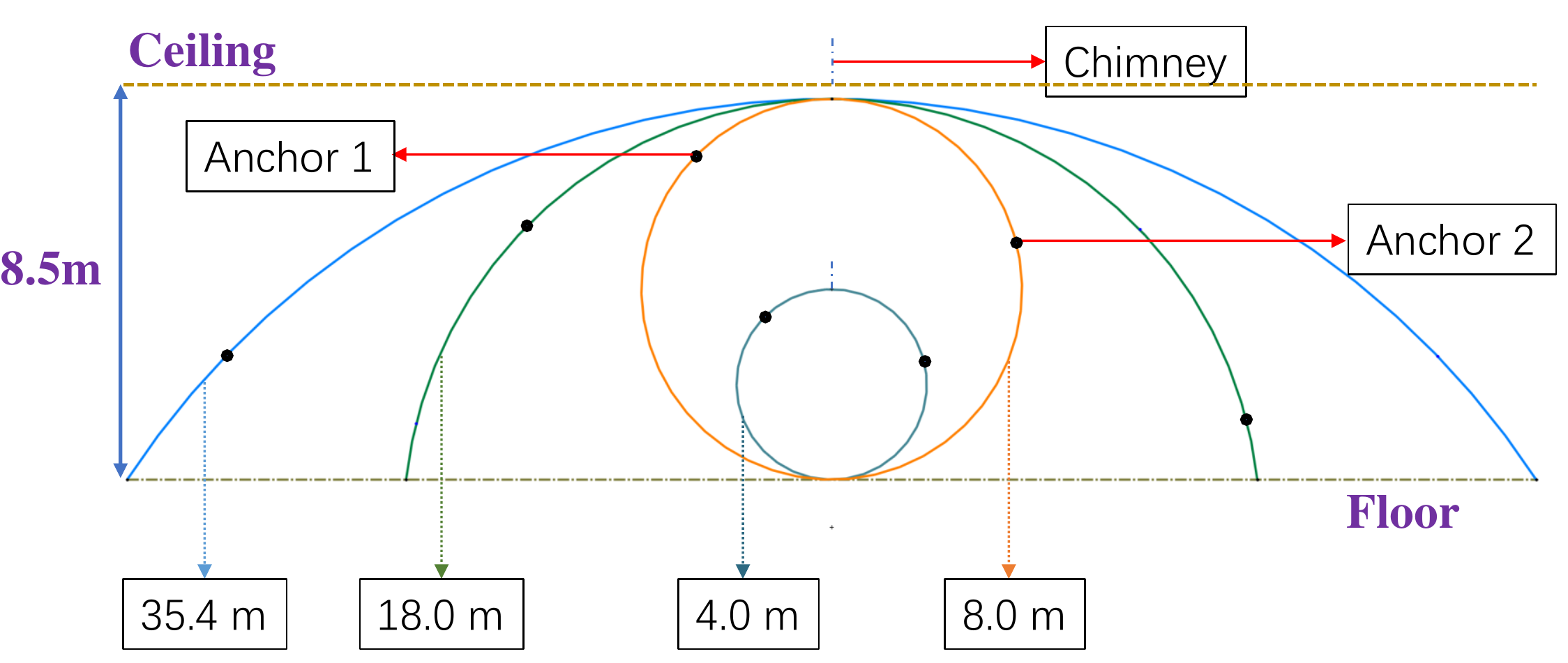}
  \caption{The mockup CLS planes constructed in the lab. Each circle represents a scaled version of the CD and corresponding CLS plane (see also Fig. 1), with the CLS anchor locations indicated as the black points.}
  \label{testplan}
\end{figure*}

As shown in Fig.~\ref{system}, a prototype CLS is constructed in a lab
with 8.5~m of overhead space, with both spools elevated by an aluminum
frame. A LabVIEW control software is developed to control the spool
motions. In this prototype, a mockup chimney collar and CLS anchors
are installed to simulate realistic situations in JUNO. Due to limited
vertical lab space, we varied the locations of PTFE anchors to
simulate the CLS performance in 1/8 (4~m), 1/4 (8~m), 1/2 (18~m), and
1/1 (35.4~m) scale, but with different fractional coverage (see
Fig.~\ref{testplan}). The source positions are surveyed with a Total
Station (KTS-442R4~\cite{total_station}) with a few~mm position
accuracy.

\section{Motion sequence}
\label{sec:calibration_strategy}

\begin{figure*}[!htbp]
  \centering
  \includegraphics[width=0.6\textwidth]{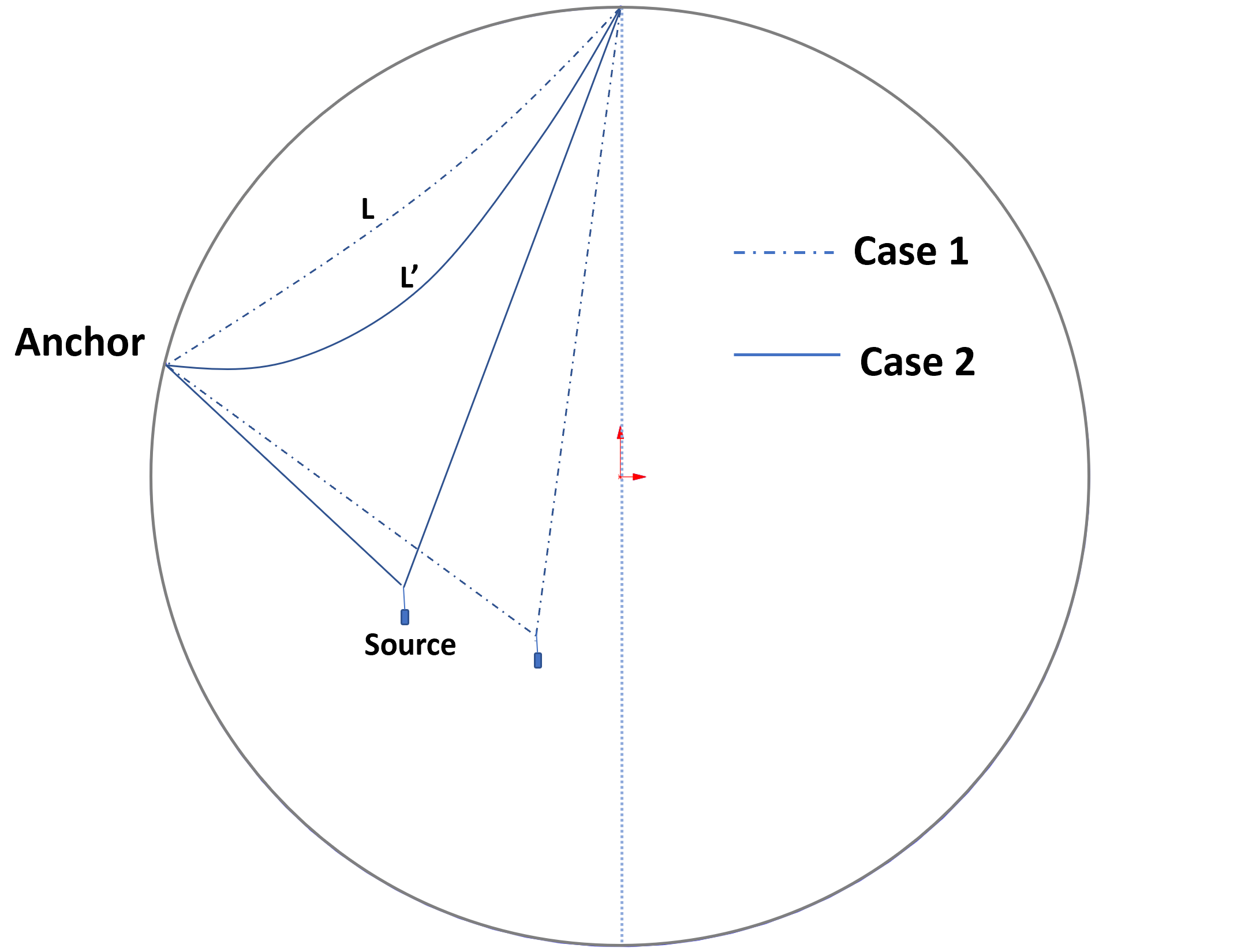}
  \caption{Illustration of the influence of the friction and cable
    self-weight to the CLS source position repeatability.}
  \label{fig:twoCase}
\end{figure*}

Realistic effects introduce difficulties in the CLS source position
control. As illustrated in Fig.~\ref{fig:twoCase}, the self weight of
the cable and the friction from the anchor to the side cable have a
significant influence to the movement of the side cable. For example,
in Case 1 and Case 2, the center cable stays the same and the side
cable is actively pulled or passively released, respectively, reaching
the same final cable length. The locations of the source are clearly
different, and in Case 2 the section L$'$ can be curved with poor
repeatability (backlash). Therefore, in our motion sequence, the side
cable should only be pulled or stay stationary. The center cable, on
the other hand, can be released to reach the designated length as it
is mostly free.

\newpage

\begin{figure}[!htb]
  \centering
  \subfloat[]
  {
    \begin{minipage}[t]{0.45\linewidth}
        \centering
        \includegraphics[width=1.0\textwidth]{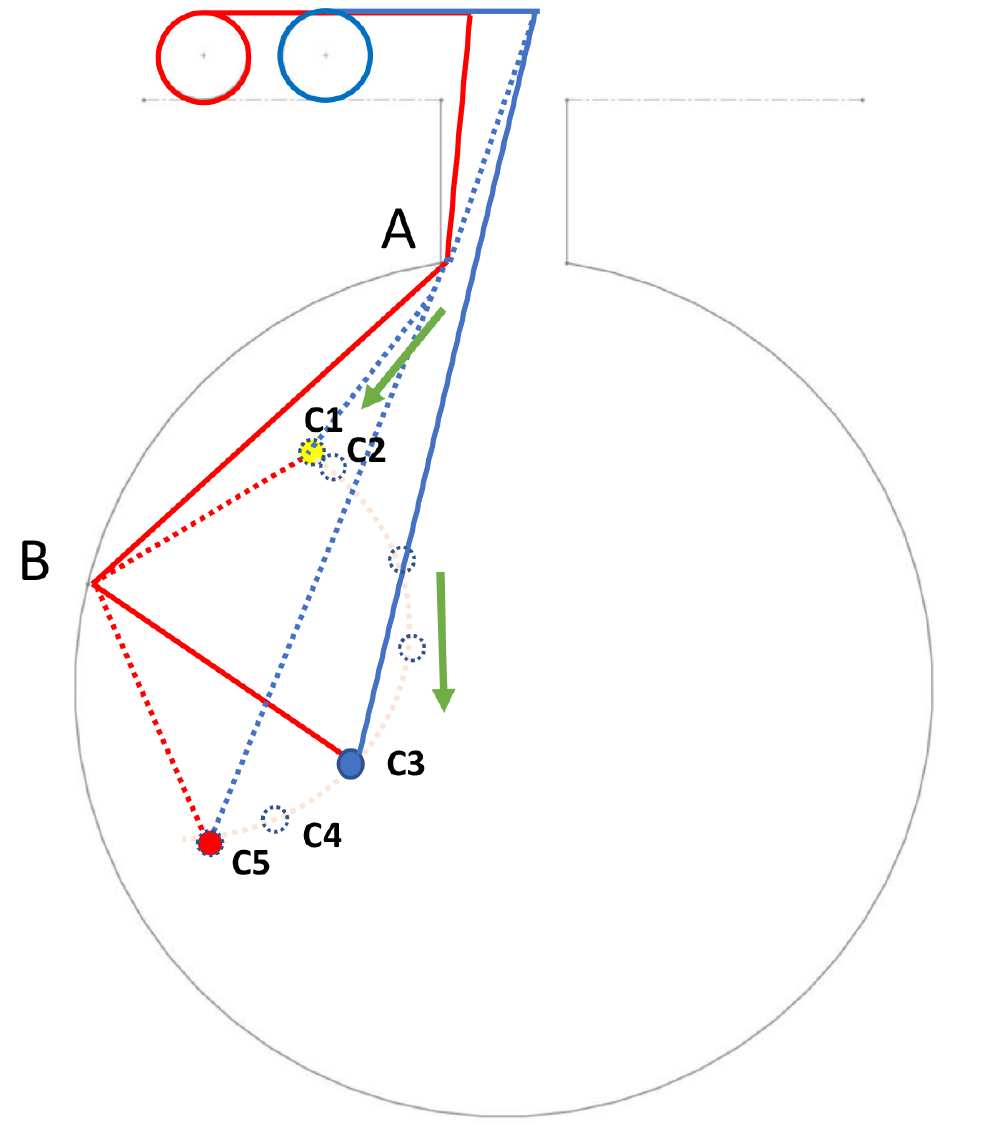}
        \label{stepIntro1}
    \end{minipage}%
  }
  \subfloat[]
  {
    \begin{minipage}[t]{0.5\linewidth}
        \centering
        \includegraphics[width=1.0\textwidth]{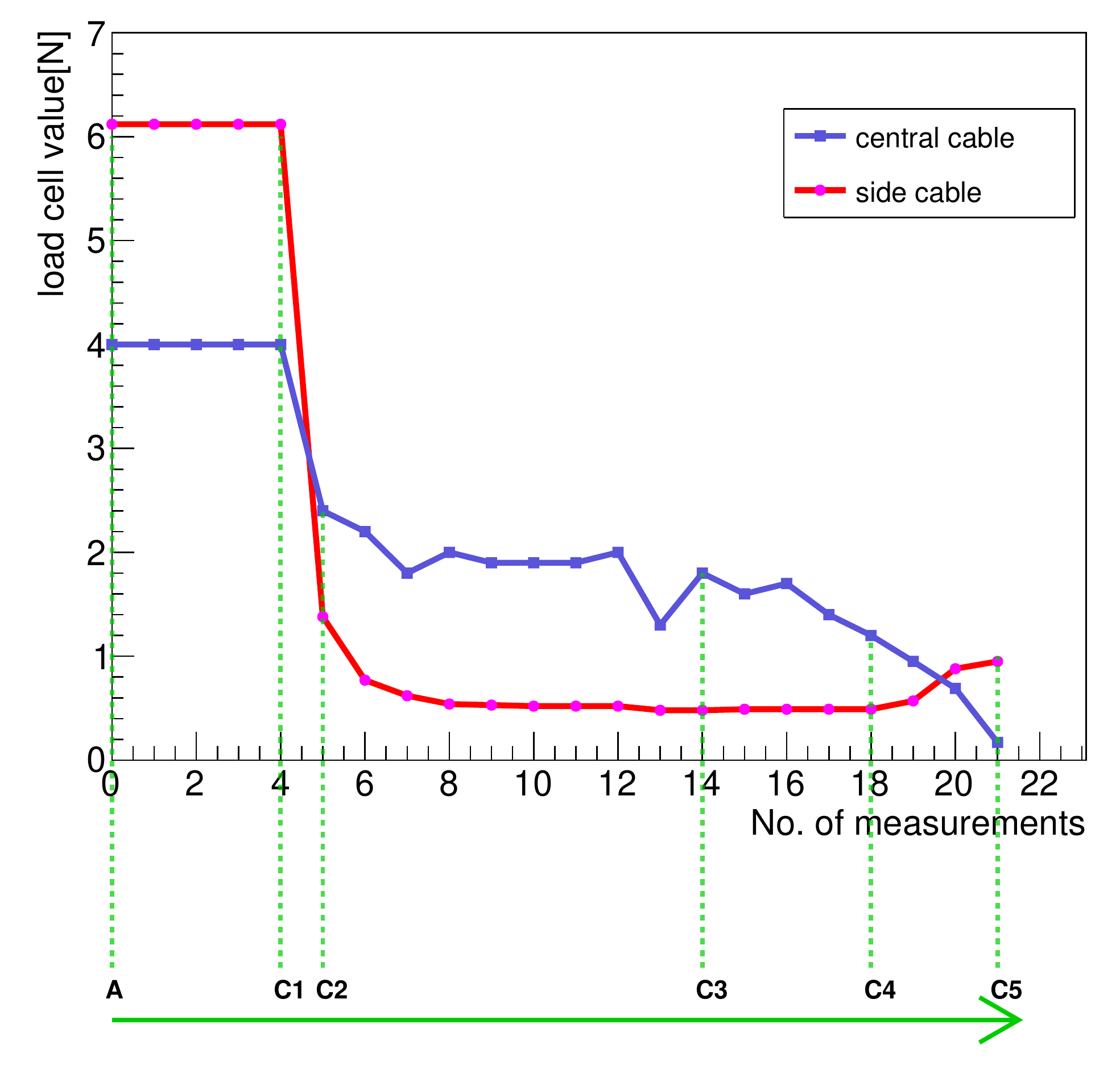}
        \label{tensionMonitoring1}
    \end{minipage}
  }%
  \caption{(a): illustration of the source motion from A to C1 to C5. Initially the central and side cables in the LS are mostly straight as we move the source through the chimney. When the source reaches the LS sphere, point C on Fig.~\ref{CLS_total_concept} is at point A. To avoid the backlash in Fig.~\ref{fig:twoCase}, it is sensible to move the source to C1 by pulling/releasing the side/central cables synchronously, then release the central cable to move the source to its designations (C2, C3, etc). (b): cable tension monitored by two load cells.}

\end{figure}

Following this guideline, the procedure of a source deployment
sequence is illustrated in Fig.~\ref{stepIntro1}. During the process, we
study the tensions inside both cables using load cells close to the
spools (Fig.~\ref{winding_mechanism}). The results are shown in
Fig.~\ref{tensionMonitoring1}. From point A to C1, the tension is
kept to be 6(4)~N in the central (side) cable, as both cables are mostly straight. From C1 to C2, there is a sharp decrease of the tensions due to sudden relaxations in both cables. The tension in the central
cable decreases gently from C2 to C5 as the gravity of the source
is taken up more and more by the side cable. However, the tension in
the side load cell does not feel this until the source passes C4
when the friction at the anchor is overcome by the gravity of the
source, leading to an increase in the load cell. Point C5 is the
lower boundary of the motion, as the central cable loses its tension
which triggers the interlock. We shall refer to this boundary as the
CLS side boundary in the remainder of this paper.

\newpage

\begin{figure}[!htb]
  \centering \subfloat[] {
    \begin{minipage}[t]{0.45\linewidth}
        \centering
        \includegraphics[width=1.0\textwidth]{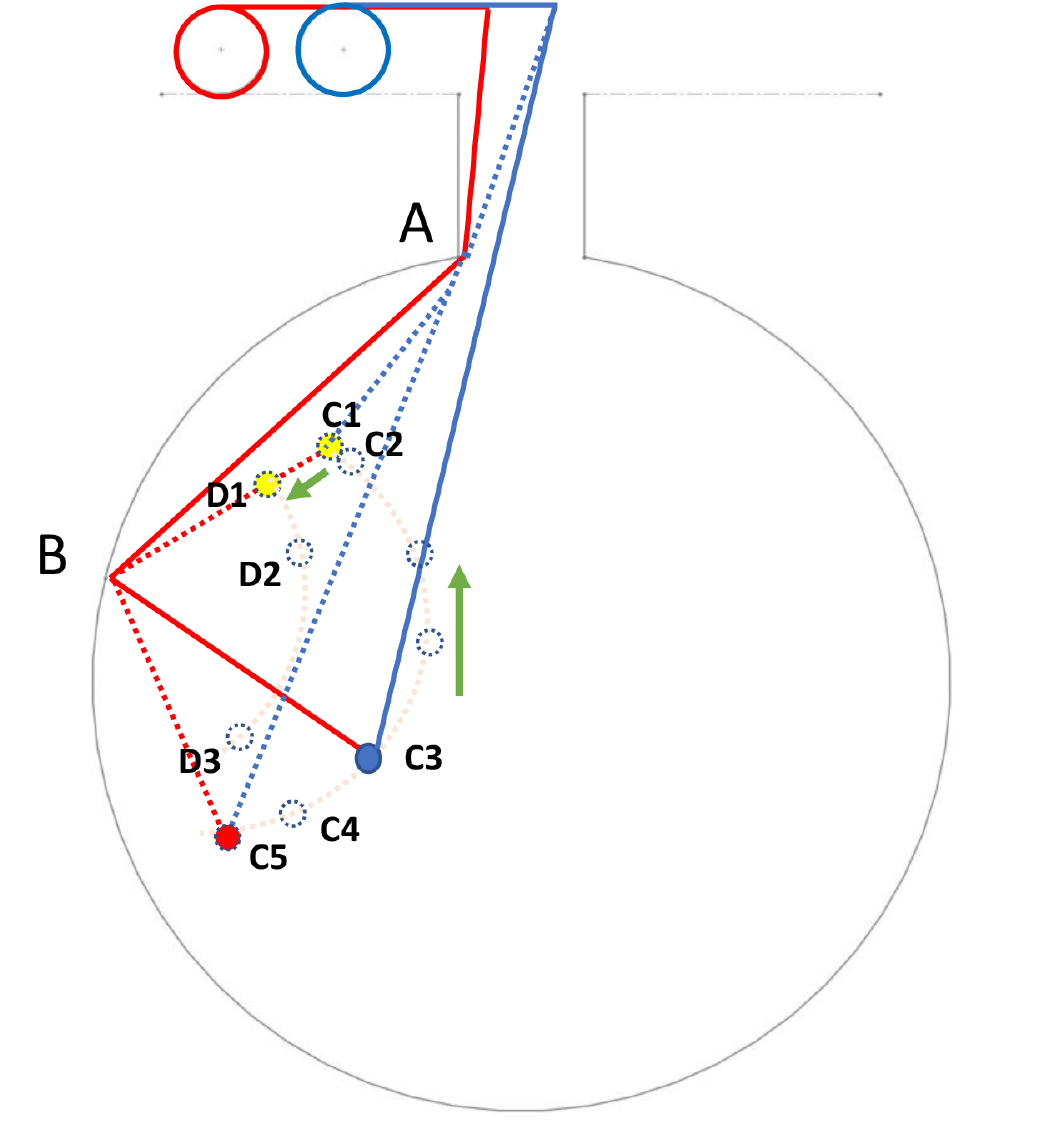}
        \label{stepIntro2}
    \end{minipage}%
  }
  \subfloat[]
  {
    \begin{minipage}[t]{0.5\linewidth}
        \centering
        \includegraphics[width=1.0\textwidth]{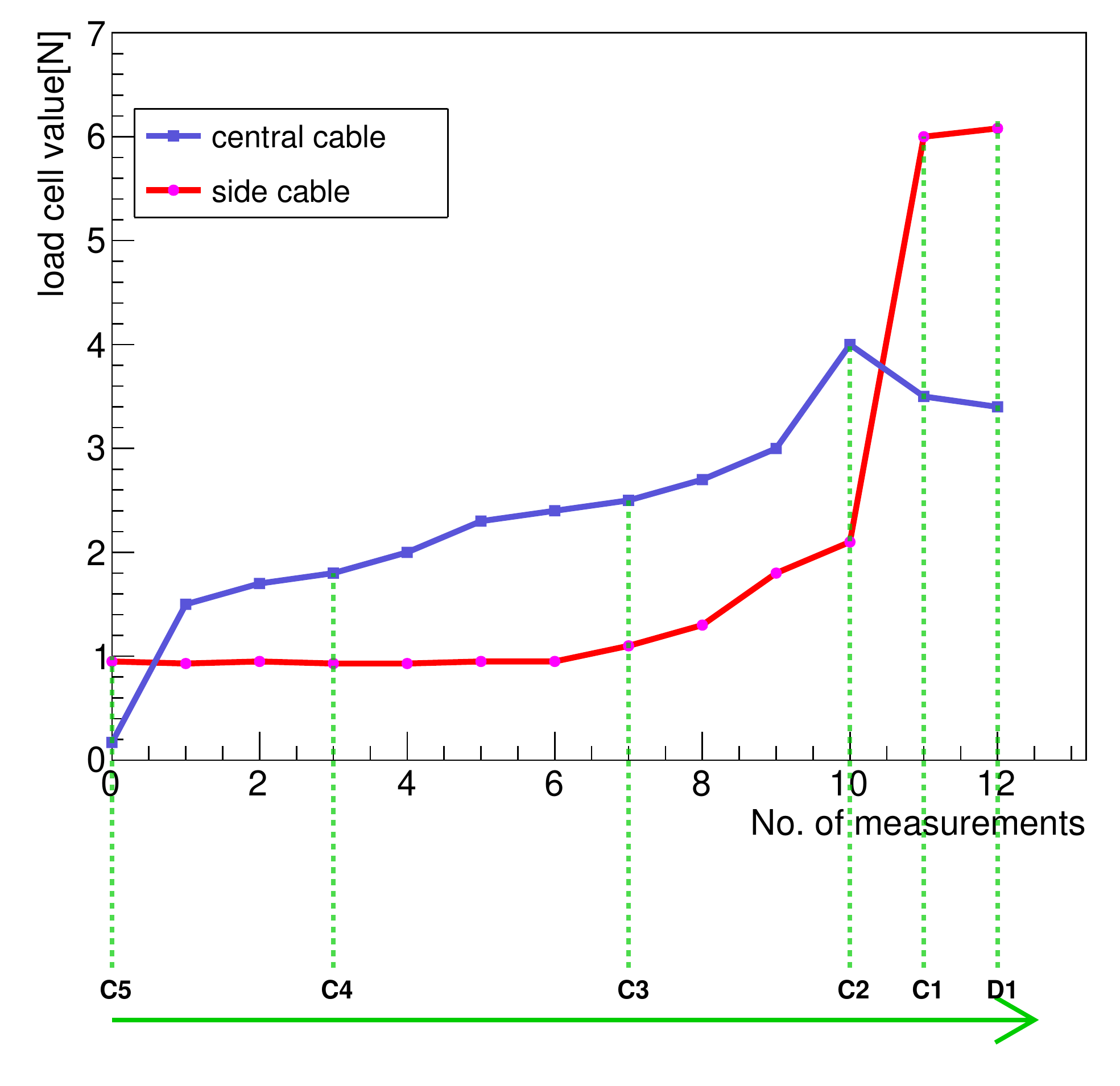}
        \label{tensionMonitoring2}
    \end{minipage}
  }%
  \caption{(a): moving from $C1$ to $D1$ by shortening the side cable, then along the $D1$ to $D3$ arc; (b): tensions in the cables.}
\end{figure}

To go to the next (shorter) side cable length, we first move the
source back from C5 to C1 by pulling the center cable, then do the
synchronous motions to D1 (Fig.~\ref{stepIntro2}). The motion along
the arc from D1 to D2 etc. can be carried out similarly. The readings of
load cell from C5 to D1 is shown in Fig.~\ref{tensionMonitoring2}. One
observes increase of tension in the central cable as it is taking up
more and more the gravity component. The tension in the side cable
increases abruptly when the source moves from C2 to C1, as segments
A--C1--B approach the straight line A--B. However, with limited
tension in the cables, C1 can never touch A--B (the upper boundary of
CLS).  Such a procedure should be repeated in a single direction
(reducing the side cable length) until the source gets close to B. To
bring the source back to home, it will be moved along B--A, by the
reverse synchronous motion (releasing the side cable and pulling the
central cable).

\newpage

In the entire process, the tension in the cables should be kept to be
larger than 0.2~N to avoid cable slipping grooves on the spool and
less than 10~N to protect the PTFE surfaces in direct contact with the
cables.

\section{System performance}

\label{sec:performance}
Using the system shown in Fig.~\ref{testplan}, realistic tests on the CLS prototype have been performed in circles with different diameters. The gravity of the source assembly is chosen to be 150~g, the effective weight after considering the LS buoyancy. To simulate the effect of the LS buoyancy to the 1~mm CLS cable (LS density $0.859$ g/cm$^{3}$), the performance of 1~mm (350~g/100~m) and 0.8~mm (220~g/100~m) cables are compared. The friction is primarily due to CLS cable in touch with the chimney collar and the CLS anchor. In our tests, the collar and the anchor are made with the final design and surface treatment. Although infiltration of the LS to the PTFE surface would reduce the friction, the mock tests are performed with dry surfaces as a worst case.

\subsection{Accessible area}
\begin{figure}[!htbp]
  \centering
  \includegraphics[width=5in]{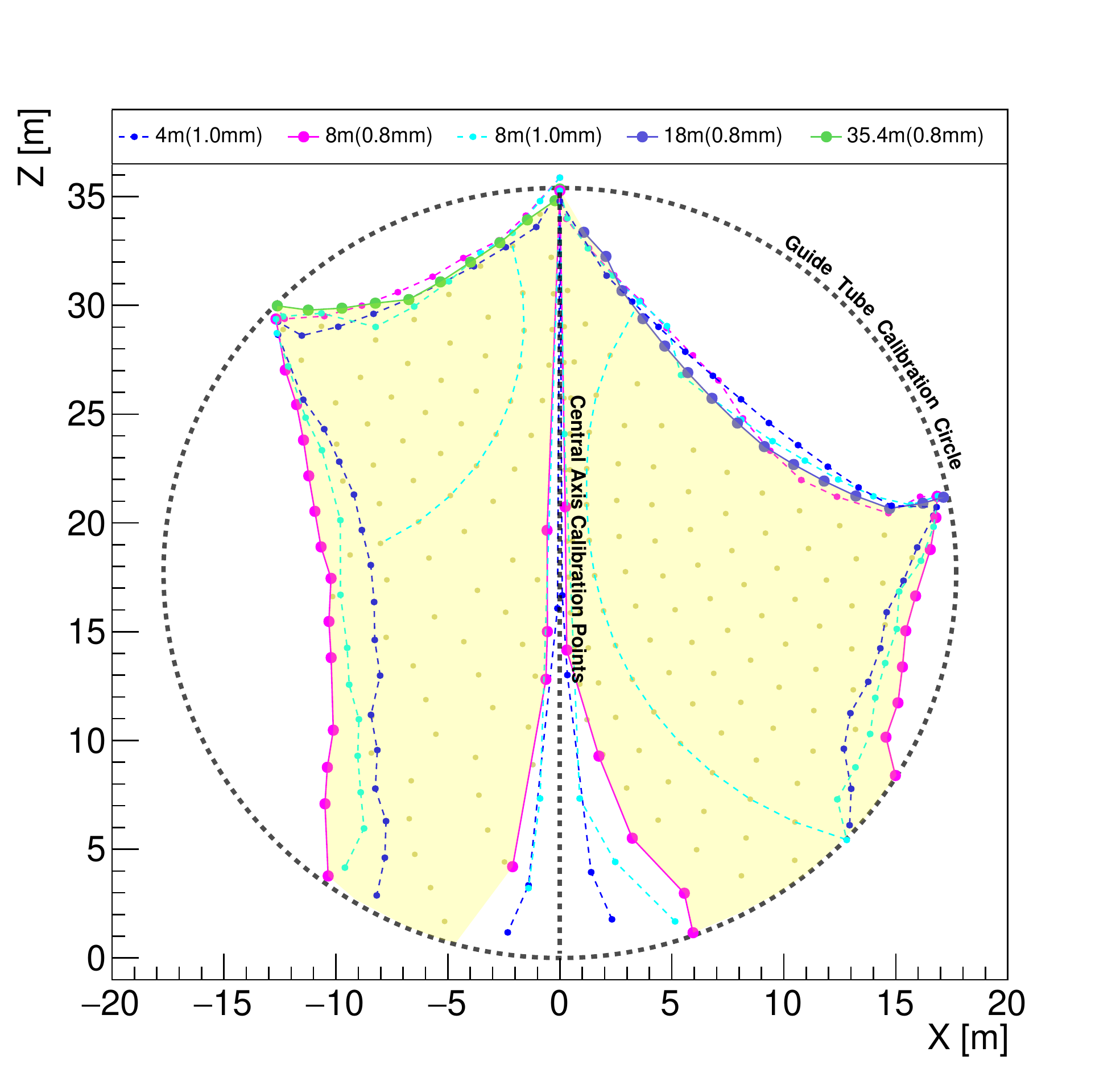}
  \caption{Results of the CLS tests. Each point in the figure
    represents a real location that was achieved in the tests, with
    different colors representing deployments in difference circles
    with a given cable (see legend). The green dashed circles
    indicate theoretical arcs when pulling the central cable with side
    cable stationary. The shaded area represents the predicted
    coverage in a full CD. JUNO also plans to have a central calibration unit covering the entire central axis(black-dashed line), and a guide tube covering a vertical circle of the sphere(black-dashed circle)~\cite{ZFY_paper}, effectively increasing the positional coverage.}
  \label{CLSresults}
\end{figure}

Tensions in the cables are monitored during the tests. The effective
coverage area is defined by three boundaries. The side boundary
(tension in the central cable hitting the lower limit) and the upper
boundary (tension in cables reaching the upper limit) have been
discussed in Sec.~\ref{sec:calibration_strategy}. The central
boundary, on the other hand, is due to the fact that non-zero tension
in the side cable and its self-weight always apply a horizontal force
to bias the location of the source from the vertical line. These
boundaries are shown in Fig.~\ref{CLSresults}.

\newpage

From the measurements, one observes that the accessible area 1) is mildly larger with the 0.8~mm cable,
as expected, and 2) the upper, side, and central boundaries are going
lower, outward, and less vertical when the diameter of the circle
increases. We predict real CD coverage based on the 0.8~mm results
with boundaries obtained in the largest possible circles in the test,
shown as the shaded region in Fig.~\ref{CLSresults}. In each half
plane, the accessible area can reach 60\% for the $46^{\circ}$ anchor,
and 66\% for the $76^{\circ}$ anchor. When the azimuthal symmetry is
considered, the entire plane can be covered to 79\%. The redundant
coverage between the two half-planes also serves as a critical
cross-check of the azimuthal-symmetry.

\subsection{Repeatability}

\begin{figure}[!htbp]
  \centering
  \includegraphics[width=5in]{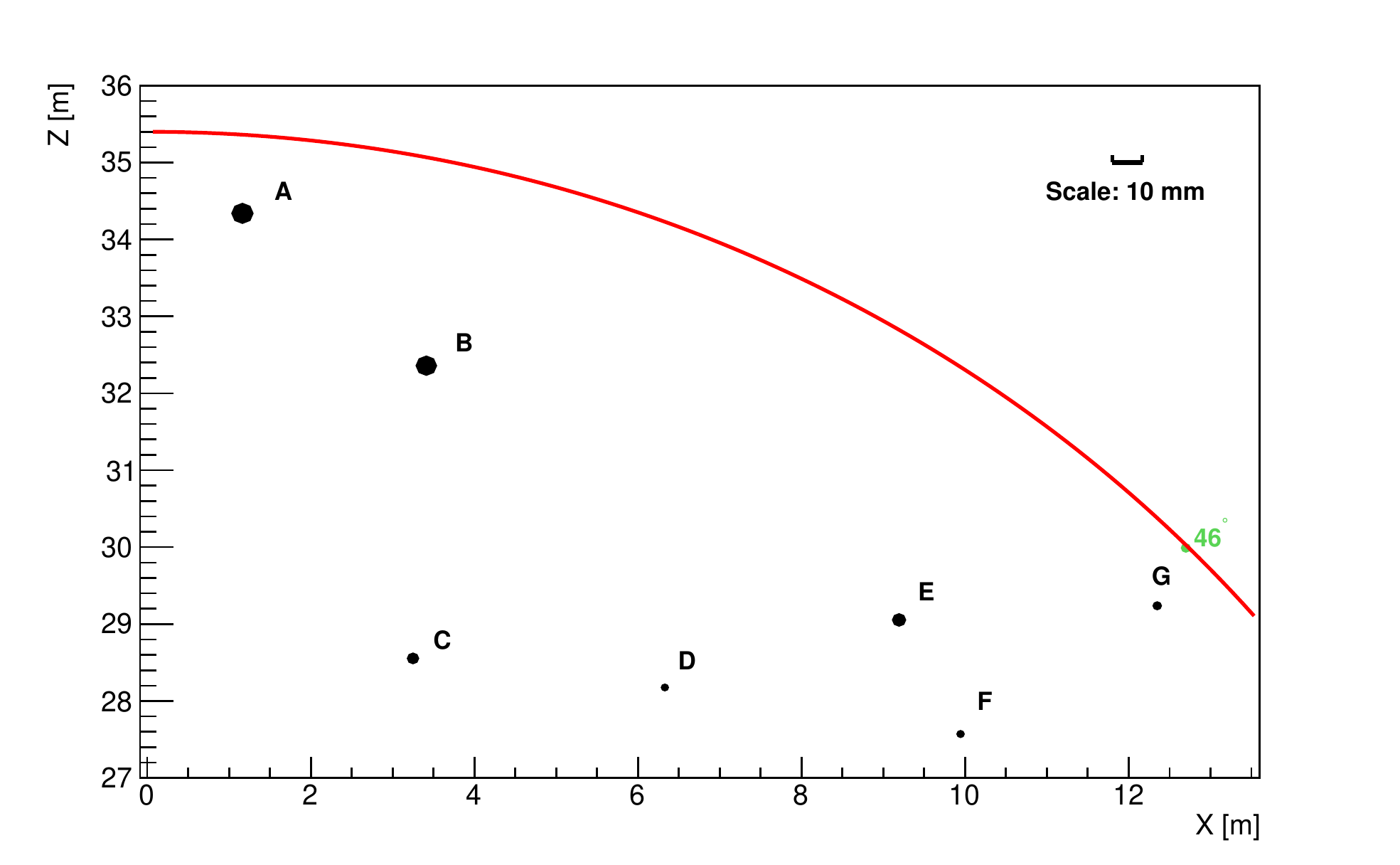}
  \caption{Position repeatability of seven points in the 35.4~m CLS
    test, with accessible locations limited by the lab setup. $Z=35$~m
    represents the chimney-sphere joint. For each point, the diameter indicates the standard deviation from the position in the first deployment in the twelve repeated deployments. The typical values are shown in Table. ~\ref{repeat}.}
  \label{repeatability}
\end{figure}

\begin{table}[!htbp]
  \centering
  \begin{tabular}{c|c|c|c|c|c|c|c}
  \hline
  points & A & B & C & D & E & F & G \\
  \hline
  Standard deviation (mm) & 6.8 & 6.5 & 3.5 & 2.2 & 4.1 & 2.1 & 2.5 \\
  \hline
\end{tabular}
  \caption{Standard deviation for seven points in the repeatability tests}\label{repeat}
\end{table}

Ideally, calibration points are uniquely determined by the lengths
of the cables~\footnote{Not by trigonometry, but by an empirical
  lookup table if an absolute calibration is carried out.}, but
frictions, self-weight, etc are expected to introduce
uncertainties. In the 35.4~m test, motion sequence of $ \rm A
\rightarrow \rm B \rightarrow \rm C \rightarrow \rm D \rightarrow \rm E
\rightarrow \rm F \rightarrow \rm G \rightarrow \rm A \rightarrow \dots$
(Fig.~\ref{repeatability}) are repeated twelve times and we measure
the repeatability of the locations of the seven points. The results are shown in
Fig.~\ref{repeatability}, with a worst standard deviation of 6.8~mm. This indicates that once a
motion sequence is fixed, the friction and backlash uncertainties are
under good control. We note that in the real experiment, the absolute
source position will be measured by the ultrasonic
system. Nevertheless, the cable lengths can serve as an importance
cross-check.

\section{Summary}
\label{sec:summary}

In this paper, we report the design and prototype performance of one
of the key calibration systems, the cable loop system, of the JUNO
experiment. The robustness of the system is tested rigorously with a
complete function prototype at smaller scales in comparison to the real experiment. Based on the measurements, the CLS is capable to cover
about 79\% to a full plane of the central detector, with \textless
10~mm positional repeatability. In combination with an extensive calibration program discussed in Ref.~\cite{ZFY_paper}, this CLS can satisfy the demanding requirements of the JUNO calibration.

\section{Acknowledgement}

This work is supported by the Strategic Priority Research Program of
the Chinese Academy of Sciences, Grant No. XDA10010800, and the CAS
Center for Excellence in Particle Physics (CCEPP). We thank the
support from the Office of Science and Technology, Shanghai Municipal
Government (Grant No. 16DZ2260200), and the support from the Key
Laboratory for Particle Physics, Astrophysics and Cosmology, Ministry
of Education.

\bibliography{mybibfile}

\end{document}